\documentclass[conf]{new-aiaa}
\usepackage[utf8]{inputenc}
\usepackage{subcaption}
\usepackage{graphicx}
\usepackage{amsmath}
\usepackage[version=4]{mhchem}
\usepackage{siunitx}
\usepackage{longtable,tabularx}
\title{Dynamic Lane Allocation in UAM Corridors for Efficient Multimodal Door-to-Door Mobility}

\author{
Andrew Park\footnote{PhD Student, Department of Civil and Environmental Engineering, University of California, Berkeley, AIAA Student Member.}, 
Jordan Kam\footnote{Undergraduate Student, Aerospace Engineering Program, University of California, Berkeley, AIAA Student Member.}, 
Vishwanath Bulusu\footnote{Lecturer, Department of Civil and Environmental Engineering, University of California, Berkeley, AIAA Member.},
Alexandre Bayen\footnote{Professor, Department of Electrical Engineering and Computer Science, University of California, Berkeley.}, 
and Raja Sengupta\footnote{Professor, Department of Civil and Environmental Engineering, University of California, Berkeley, AIAA Member.}}
\affil{University of California, Berkeley, Berkeley, CA, 94720}

\begin{document}

\maketitle

\begin{abstract}
This article presents dynamic directional lane allocation in urban air mobility (UAM) corridors as a discrete-time mixed-integer linear program (MILP). This formulation activates, deactivates, and reverses lane direction as bi-directional airspace demand evolves. We model demand from disaggregate ground travel data by decomposing each trip into a multi-modal sequence with first-, middle-, and last-mile legs and routing the UAM-served middle-mile segment through a vertiport-side dispatch model. We use the San Francisco Bay Area as a case study by placing a multi-region spanning corridor between Contra Costa county and Silicon Valley. We find that the dynamic policy cuts unused airspace capacity by $5\times$, increases mean lane utilization from 36--48\% to 67\% at the same service level relative to baselines, and reduces commuting-population mean travel time by up to 21.6\%. These results show that dynamic configuration of airspace capacity alleviates a significant percentage of the under-utilization issue of lane-based UAM airspace design and UAM concept of operations. This dynamic allocation also provides a safe, structural way to increase throughput, making UAM a more viable complement to multimodal door-to-door mobility systems.
\end{abstract}


\section{Nomenclature}

{\renewcommand\arraystretch{1.0}
\noindent\begin{longtable*}{@{}l @{\quad=\quad} l@{}}
$a_{ijt}$ & lanes activated into the $i \to j$ direction at time $t$ \\
$C_s$ & penalty weight for operational friction (switching cost) \\
$C_u$ & penalty weight for unserved demand \\
$C_w$ & penalty weight for wasted capacity \\
$E$ & set of bi-directional corridors \\
$F_{ijt}$ & forecasted flight demand from node $i$ to node $j$ at time $t$ \\
$i, j$ & node indices representing corridor endpoints \\
$K$ & throughput capacity of a single active lane per time step \\
$L_{ij}$ & total physical lanes available between node $i$ and node $j$ \\
$s_{ijt}$ & shortfall (rejected flight slots) from $i$ to $j$ at time $t$ \\
$T$ & set of time steps over the planning horizon \\
$t$ & time step index \\
$\Delta t$ & duration of one time step \\
$v_{ijt}$ & lanes deactivated from the $i \to j$ direction at time $t$ \\
$w_{ijt}$ & wasted capacity (unused flight slots) from $i$ to $j$ at time $t$ \\
$y_{ijt}$ & active, usable lanes operating from $i$ to $j$ at time $t$ \\
$Z$ & total system penalty (objective function value) \\
$\tau$ & flush time (in time steps) required to clear a lane \\
$p_{c}$ & market share of UAM
\end{longtable*}}

\section{Introduction}

Door-to-door mobility in densely populated metropolitan regions is often considered as a multimodal transportation problem~\cite{di2022outline}. Residents of outlying or underserved communities of major cities face commute times that scale poorly with distance~\cite{niedzielski2014travel}. Ground road networks become highly congested at peak commute hours, and further geographic separation between residential clusters and employment centers makes high-frequency fixed-route transit difficult to implement. Advanced air mobility (AAM) is a promising mode of transportation that may help relieve some of these effects by creating faster commutes and transit times in regions previously considered difficult to connect~\cite{goyal2018urban, kam2025operational}. Such difficulties come from specific characteristics of traditional transportation modes. For instance, heavy and light rail offer high throughput on dense corridors but require capital investment that limits coverage to the highest-volume routes; bus networks extend coverage but are constrained by the same congested road network they share with private vehicles; private cars and ride-hailing offer door-to-door flexibility at the cost of road occupancy, emissions, and travel times that grow superlinearly during rush hours. AAM and its near-term subset of urban air mobility (UAM) have been proposed as a complement rather than a substitute \cite{idris2024passenger}. These services can compress middle-mile travel between regions while relying on ground-based transportation for first- and last-mile access to dispersed origins and destinations. Hence, AAM does not replace existing transit systems but fills the gap that ground transit is inefficiently serving today.


However, AAM must maintain sufficient middle-mile throughput to make a meaningful contribution to people's daily transportation plans. At a significant adoption level of AAM, its throughput is constrained by airspace~\cite{vascik2018scaling}. The low-altitude airspace where AAM is intended to operate is already shared with general aviation, helicopter, and drone traffic, and the Federal Aviation Administration (FAA) is not planning to provide air traffic control (ATC) services to UAM aircraft within UAM Corridors. Cooperative airspace management, performed by the federated network of Providers of Services for UAM (PSUs) designated by the FAA UAM Concept of Operations (ConOps)~\cite{faa_uam_conops_v2_2023}, is therefore the operational layer that determines whether AAM scales as a multimodal complement. The ConOps assigns PSUs the responsibility for conflict management, equity of airspace usage, and Demand-Capacity Balancing (DCB) on behalf of UAM operators, and leaves the specific cooperative practices that implement these responsibilities as Cooperative Operating Practices (COPs). These are industry-defined, FAA-approved methods that the regulatory framework establishes the need for but does not itself prescribe~\cite{faa_uam_conops_v2_2023, lee_x5_2025}. The COP white space is where novel airspace design and operations, methods for more efficient airspace use, and the algorithmic primitives that a future ATC for AAM will require are studied \cite{su2025flight}.


Within this space, multiple airspace structure designs have been proposed over the years, including free flight, corridor, tube, and lane~\cite{Yang_airspaceReview2024}. The lane system, while the most safety-critical, has been criticized as the most inefficient form of airspace usage. This criticism, however, reflects the cost of static lane assignment under asymmetric demand; lane structure itself supports tighter aircraft spacing than less constrained structures \cite{sacharny2020faa, henderson2024multi}, and we argue that dynamic directional reallocation captures this density advantage while maintaining high airspace utilization rate. In the literature, this is a form of supply-side control of UAM operations. However, recent work has addressed congestion through demand-side controls, including pre-departure ground delays~\cite{lee_x5_2025, sacherny_laneBased2022}, mission-aware rerouting~\cite{Xie_hybrid_DCB_2021, Xie_AIRerouting_2023, du2025_collaborativeFlowOptimization}, and bi-level scheduling protocols across the federated PSU architecture~\cite{wu2025optimizationguidedexplorationadvancedair, KimLi_centralDistributed2026}. The supply-side question of how the airspace should be dynamically configured to absorb time-varying directional demand has received less attention, and the work that does address it tends to bundle supply-side decisions together with scheduling-side decisions whose authority is already exercised by PSUs~\cite{Hearn_2023_dynamic_airspace_config, tang_OnDemandDCM_2022, Tang_decentralized_DCM}. The dynamic airspace reconfiguration problem for a UAM corridor with multiple lanes capable of bidirectional flow has not yet been explored.

This study explores the idea of dynamic lane allocation/de-allocation and reversal of direction as bi-directional demand evolves on a typical weekday, subject to the operational constraint that a lane's direction switching requires a buffer time to flush out remaining aircraft in the airspace before opposite flow can be admitted. We formulate this as a discrete-time mixed-integer linear program (MILP), and solve the airspace planning problem from the perspective of airspace managers to improve airspace utilization and throughput. In doing so, we show that dynamic reconfiguration mitigates the disadvantages of lane-based airspace and significantly improves passenger throughput and airspace-utilization efficiency. The remainder of the paper is organized as follows. Section~\ref{sec:methodology} presents the formulation. Section~\ref{sec:experimentation} details the experimental setup and demand sources for our Bay Area case study. Section~\ref{sec:results} reports comparative and sensitivity results, followed by a conclusion with limitations of this study and future work.


\section{Methodology}
\label{sec:methodology}

\subsection{Demand from Multi-Modal Transition}
\label{subsec:demand}

To compute the airspace demand series $\{F_{ijt}\}$ that the MILP consumes, we proceed in three steps: 1) a service-design assumption that locates UAM within a multimodal trip, 2) a market-share assumption that selects which travelers use UAM, and 3) a vertiport-side dispatch model that converts per-passenger trips into per-slot aircraft demand. We model UAM service as the middle-mile (MM) leg of a door-to-door trip, with first- and last-mile (FM/LM) legs handled by ride-hailing or robotaxi services. This three-stage decomposition is standard in UAM demand modeling~\cite{cao6521194joint} and reflects the geographic mismatch between a sparse vertiport network and dispersed trip origins and destinations. 

Treating UAM as the MM leg lets us predict demand patterns directly from existing ground travel records, which already encode the origin, destination, time, and trip-level travel times needed to evaluate UAM as an alternative for each trip. For the total addressable market, we use Replica as our primary source of ground travel demand~\cite{Replica2026}. Replica generates a disaggregate synthetic population and trip diary by combining anonymized mobile location data with census, employment, and survey data, calibrated against ground-truth mobility patterns. For our case study, the data is sourced from disaggregate trips on a single weekday in the candidate UAM service region.

A standard next step would estimate UAM market share via a discrete choice model (DCM)~\cite{train_discrete_2009}, which predicts mode shift as a function of travel time, cost, comfort, and other attributes. While methodologically proven, DCM requires complex modeling of human decision-making and dedicated data collection campaigns, and the resulting estimates carry large error bars in the case of UAM~\cite{fu_exploring_2019, ilahi_exploring_2020}. From the perspective of an airspace manager, however, the upstream mode-choice mechanism need not be resolved exactly: only its output, the realized airspace demand, needs to be specified. We therefore treat UAM market share as a fixed parameter, the market capture rate $p_c$ (commonly referred to as market share denoted as $\theta_\text{UAM}$ in the literature), and sweep it across a range to learn study how varying UAM adoption levels affect the airspace management policy.

Given the disaggregate ground trips and a fixed capture rate $p_c$, the conversion to airspace demand $\{F_{ijt}\}$ proceeds in two stages. First, we assume that people's willingness to switch mode primarily comes from the travel time advantage, and implement a capture-rate filter that retains the top $p_c$ fraction of passengers ranked by their UAM travel-time advantage over driving (computed from Replica's mode-specific travel-time fields). Second, we adopt a rule-based dispatch policy that simulates vertiport boarding queues per OD pair, modeling how UAM operators would dispatch under realistic pooling and wait-time constraints. Passengers accumulate at their origin vertiport, and once a queue reaches $\mathrm{MIN\_LOAD}$, we dispatch as many fully-loaded aircraft of capacity $\mathrm{CAP}$ as the queue allows. Passengers who wait longer than $W$ slots without filling an aircraft defect to driving and are dropped from corridor-side demand. This is referred to as demand spill. Similar rule-based dispatch policies have been adopted in simulation studies of on-demand UAM services~\cite{onat5072017urban}. The resulting dispatched-aircraft counts, shifted in time to align with their entry into the high-throughput UAM corridor where the dynamic lane system is implemented, constitute the exogenous demand
$\{F_{ijt}\}$ that the MILP consumes. The series is computed in advance at the operator's planning cadence.




\begin{figure}[h!]
    \centering
    \begin{subfigure}{0.4\textwidth}
        \includegraphics[width=\textwidth]{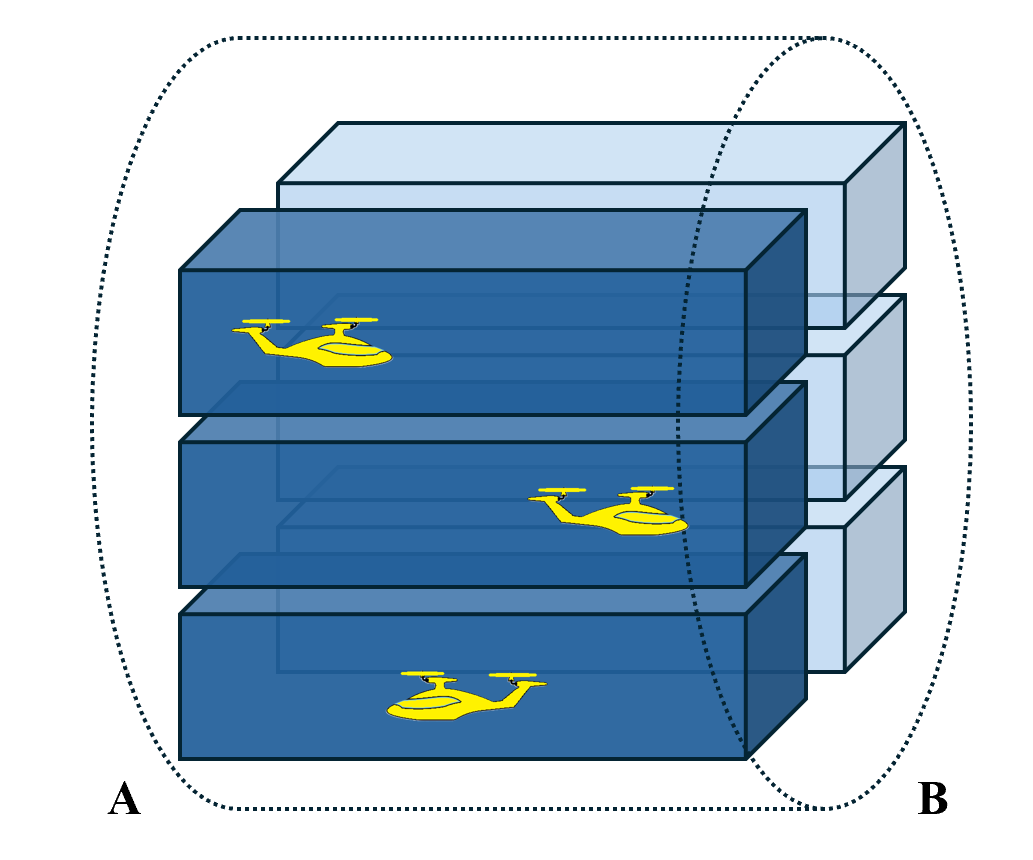}
        \caption{6-lane UAM corridor system}
        \label{fig:skylane}
    \end{subfigure}
    \hspace{0.05\textwidth}
    \begin{subfigure}{0.5\textwidth}
        \includegraphics[width=\textwidth]{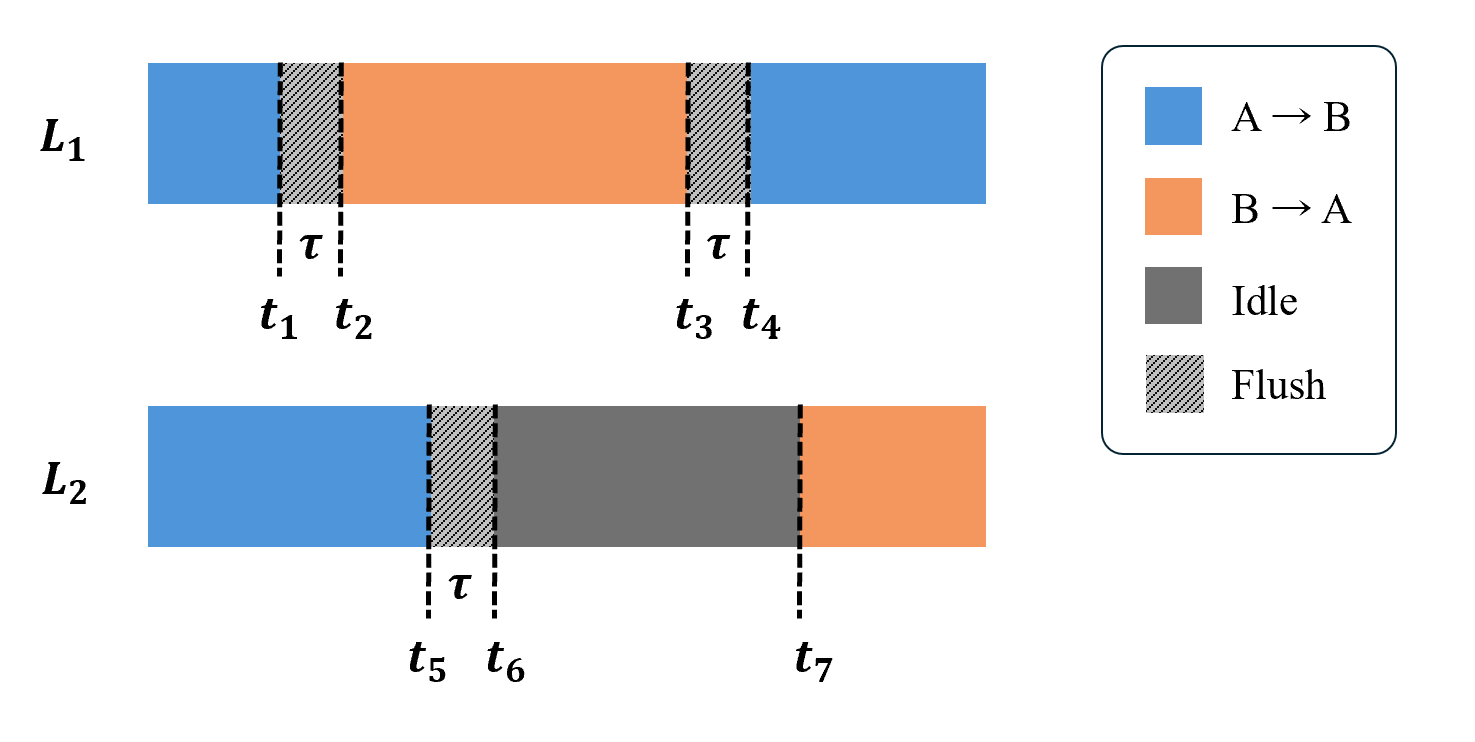}
        \caption{Visualization of lane direction-switching mechanism}
        \label{fig:dsr}
    \end{subfigure}
    \caption{(a) shows a visualization of a 6-lane corridor system that has layered lanes in a confined space. (b) illustrates the lane direction switching mechanism of UAM corridors. $L_1$ switches from $A \rightarrow B$ to $B \rightarrow A$ direction starting at $t_1$, after taking $\tau$ to flush the lane of any pre-existing aircraft in the lane, which begins its new direction at $t_2$. $L_2$ switches from $A\rightarrow B$ to an idle lane at $t_5$, after taking $\tau$ to flush the lane. At $t_7$, lane $L_2$ can immediately be allocated to serve $B \rightarrow A$ direction aircraft flow, since it was switching from an idle state.}
    \label{fig:corridor}
\end{figure}

\subsection{Concept of Operations for Dynamic Lane System in UAM Corridors}

In a dynamic lane system, the airspace manager has visibility into expected daily flow patterns within a UAM corridor and configures the corridor's lane allocation to match the time-varying directional demand. A representative stacked corridor design is illustrated in Fig.~\ref{fig:corridor}(a): the corridor connecting vertiports $A$ and $B$ is structured as $N$ parallel lanes, each operated as an independent directional channel at a distinct altitude. This layered design is consistent with the FAA UAM ConOps, which envisions corridors with multiple internal tracks to support increasing flight frequencies, and follows the pattern explored across the lane-based and grid-based UAM literature~\cite{sacherny_laneBased2022, pang_airspaceConfig2020, sunil2015metropolis}. The vertical separation $\Delta z$ between adjacent lanes is a corridor performance parameter rather than a quantity our formulation determines: the FAA does not currently specify intra-corridor vertical separation for UAM, and published research proposals span an order of magnitude depending on assumed navigation accuracy and target safety levels --- approximately 11~m for RNP10-equipped multi-rotor eVTOL at a $10^{-6}$ target level of safety~\cite{wang2024safety}, 50~m in the Metropolis layered airspace concept~\cite{sunil2015metropolis}, and up to roughly 137~m (450~ft) in the survey of Bauranov and Rakas~\cite{bauranov2021designing}. Our optimization treats each lane as an abstract directional capacity unit indexed by $l \in \{1, \dots, N\}$, and is independent of the specific geometric separation chosen so long as adjacent lanes remain operationally independent. In this work, we do not distinguish corridor from lane systems; rather, we model the lane system as a partition of the UAM corridor.

Direction switching is governed by a flush time $\tau$. Any direction change must be preceded by a $\tau$-slot interval during which no aircraft may enter the lane, allowing residual traffic to clear before the opposite flow is admitted. Figure~\ref{fig:corridor}(b) illustrates the two transition types this constraint produces. $L_1$ undergoes an active-to-active reversal that requires a flush before the new direction begins. $L_2$ deactivates to an idle state and is later reactivated in the opposite direction. The deactivation requires a flush to clear residual traffic, but reactivation from idle does not, since an idle lane already contains no aircraft.

The lane-based structure is one of several proposed for UAM airspace. Yang et al.~\cite{Yang_airspaceReview2024} survey four candidates (e.g,free flight, corridor, tube, and lane) and characterize the lane model as the most restrictive. It yields the lowest operational risk (the single-direction confined geometry eliminates most encounter classes) but also the worst static airspace utilization. Within the lane family, prior work has typically treated directional assignment as a static design parameter~\cite{pang_airspaceConfig2020} or studied scheduling decisions over a fixed lane network~\cite{sacherny_laneBased2022}; the dynamic configuration of lane direction itself remains an open question. This static-utilization criticism is particularly important under cyclic demand patterns expected in UAM service, where a lane network optimal for the morning peak may be suboptimal for the evening peak by construction.

Our formulation directly addresses this criticism. By dynamically reallocating lane directions over the planning horizon, the corridor matches its directional capacity supply to the time-varying directional demand. Lanes that would sit underutilized under a static assignment can be flipped to the dominant direction during peak imbalance, or temporarily deallocated and released to the idle pool. Released lanes carry no operational waste penalty and are available for reassignment to other purposes (e.g., cargo flights, emergency response, or adjacent corridors under demand surge) without imposing operational overhead on the active corridor. Note that this configuration-layer optimization is complementary to scheduling-layer work. Our formulation produces the network that LBSD \cite{sacherny_laneBased2022} or NASA's S\&S service would subsequently operate within.



\subsection{Dynamic Lane System as a Supply Control Problem}
We formulate intra-corridor lane allocation as an instance of Dynamic Airspace Configuration (DAC), a problem class introduced for en-route ATM by Kopardekar et al.~\cite{kopardekar2007initial} in which airspace structure is treated as a time-indexed decision variable that adapts to shifting demand under a switching cost. Conventional DAC partitions controlled airspace into sectors and reconfigures those partitions to balance air traffic controller workload, with methodological contributions spanning integer-programming approaches to sector combining~\cite{bloem2009combining}, traffic-mass-based partitioning~\cite{klein2005efficient}, and machine-learning workload prediction driving configuration choice~\cite{gianazza2010forecasting}. The UAM setting preserves the same structural template. However, UAM corridors lie outside ATC service, so the optimization objective shifts from balancing human workload to matching directional capacity supply against directional demand. Hearn et al.~\cite{Hearn_2023_dynamic_airspace_config} adapt the template by allocating free-flight versus corridor service areas across spatial partitions, and Tang et al.~\cite{tang_OnDemandDCM_2022, Tang_decentralized_DCM} embed configuration choice within a joint dynamic capacity management MILP over slot allocation, rerouting, and pre-defined configurations. Our formulation is in the same family as Tang's but isolates the supply-side decision at finer granularity. We adopt per-lane direction within a single corridor, rather than selection from a menu of combinatorial cell-merge patterns so that the resulting COP deploys at the PSU layer without requiring authority over operator-side flight planning, and admits an explicit flush-time primitive governing safe lane reversal that has no analog in either the partitional decisions of conventional DAC or the configuration-switching of Tang's formulation.

\subsubsection{Sets and Parameters}
We index time by discrete steps $t \in T$ of duration $\Delta t$ over the planning horizon and consider a single bi-directional corridor connecting nodes $i$ and $j$ with $L_{ij}$ physical lanes shared across the two directions; the formulation extends to a network $E$ of corridors by replication with no cross-corridor coupling, so we suppress the corridor index in what follows. The exogenous demand series $F_{ijt}$ and $F_{jit}$, expressed as aircraft per slot per direction, are treated as inputs to the optimization. Each active lane provides a per-slot throughput of $K$ aircraft.

The configuration state at time $t$ is the pair of integer lane
counts $(y_{ijt}, y_{jit})$ giving the number of active lanes in
each direction; the residual $L_{ij} - y_{ijt} - y_{jit}$ accounts for lanes that are either fully idle or currently traversing a flush window. Configuration changes occur through two non-negative integer event variables: $v_{ijt}$, the number of lanes deactivated from direction $i \to j$ at time $t$ (each entering a $\tau$-slot flush before becoming idle), and $a_{ijt}$, the number of previously idle lanes activated into direction $i \to j$ at time $t$ (which require no flush, since an idle lane already contains no aircraft). Two auxiliary continuous deviational variables, $s_{ijt}$ and $w_{ijt}$, record the per-slot mismatch between provided capacity and realized demand: shortfall when demand exceeds capacity, and waste when capacity exceeds demand. The reverse-direction variables $y_{jit}, v_{jit}, a_{jit}, s_{jit}, w_{jit}$ are defined symmetrically.

\subsubsection{Objective Function}
The optimization minimizes a weighted sum of three operational
penalties --- unserved demand, switching friction, and wasted
capacity --- across both directions and the full horizon:
\begin{equation}
\label{eq:objective}
\min Z = \sum_{t \in T} \left[\, C_u (s_{ijt} + s_{jit})
                              + C_s (v_{ijt} + v_{jit})
                              + C_w (w_{ijt} + w_{jit}) \,\right].
\end{equation}

The relative weights $(C_u,\, C_s,\, C_w)$ encode the operator's
priority among service level, operational stability, and supply-side efficiency. The switching cost is asymmetric by design: only deactivations $v$ enter the objective, because a deactivation triggers the $\tau$-slot flush during which the lane is neither serving the current direction nor yet available for the opposite. Activations from idle lanes are free, since an idle lane requires no clearing before assignment.

\subsubsection{Constraints}
The corridor's physical capacity constraint must be satisfied at every time step. The active lanes in both directions, together with all lanes still traversing a flush window from a recent deactivation, cannot exceed the corridor's lane count:
\begin{equation}
\label{eq:physical_capacity}
y_{ijt} + y_{jit} + \sum_{k = t - \tau + 1}^{t}
                    \left( v_{ijk} + v_{jik} \right) \leq L_{ij}.
\end{equation}
The summation encodes the operational cost of a direction switch directly into the capacity budget. A lane in flush counts against the lane count but contributes no throughput, ensuring the formulation cannot collapse into the trivial myopic policy that flips lane direction at every slot. We express Eq.~\eqref{eq:physical_capacity} as an inequality so that the slack between its two sides represents a pool of fully idle lanes, neither active nor flushing, that the operator may release to other uses . To make the summation well-defined at the start of the horizon, we treat $y_{ij,0}$ and the historical deactivation series $\{v_{ijk}\}_{k \leq 0}$ as known constants reflecting the corridor's state when the planning window opens.

The lane state evolves between successive slots according to the event variables. The active lane count in each direction equals the previous slot's count, less any lanes deactivated, plus any idle lanes newly assigned to that direction. 
\begin{equation}
\label{eq:lane_evolution}
y_{ijt} = y_{ij,\, t-1} - v_{ijt} + a_{ijt}.
\end{equation}

Finally, the deviational variables couple configuration to demand. Capacity below demand produces shortfall and capacity above demand produces wasted capacity:
\begin{equation}
\label{eq:deviation_ineq}
s_{ijt} \geq F_{ijt} - K \cdot y_{ijt},
\qquad
w_{ijt} \geq K \cdot y_{ijt} - F_{ijt},
\qquad
s_{ijt},\, w_{ijt} \geq 0.
\end{equation}
Because Eq.~\eqref{eq:objective} penalizes both with positive weights, the optimizer drives at most one of the pair to a strictly positive value in any slot --- they behave as the deviational variables of a goal-programming formulation. We can therefore replace the pair of inequalities with a single equality without altering the optimal solution:
\begin{equation}
\label{eq:goal_programming}
K \cdot y_{ijt} - F_{ijt} = w_{ijt} - s_{ijt}.
\end{equation}
When provided capacity exceeds demand the right-hand side absorbs the excess as $w_{ijt}$; when capacity falls short it absorbs the shortfall as $s_{ijt}$. The same equality applies to the reverse direction. This collapse from two inequalities to one equality per direction halves the demand-side constraint count in the model, a non-trivial saving when the formulation is replicated across many corridors and time steps.

The complete formulation comprises the objective Eq.~\eqref{eq:objective} subject to the physical-capacity constraint Eq.~\eqref{eq:physical_capacity}, the state-evolution constraint Eq.~\eqref{eq:lane_evolution}, and the goal-programming balance Eq.~\eqref{eq:goal_programming}, with $y_{ijt}, v_{ijt}, a_{ijt} \in \mathbb{Z}_{\geq 0}$ and $s_{ijt}, w_{ijt} \in \mathbb{R}_{\geq 0}$. The result is a mixed-integer linear program with $\mathcal{O}(|T| \cdot |E|)$ variables and constraints, well within the reach of standard solvers at the urban-corridor scale.

\begin{figure}[t!]
\centering
\includegraphics[width=.6\textwidth]{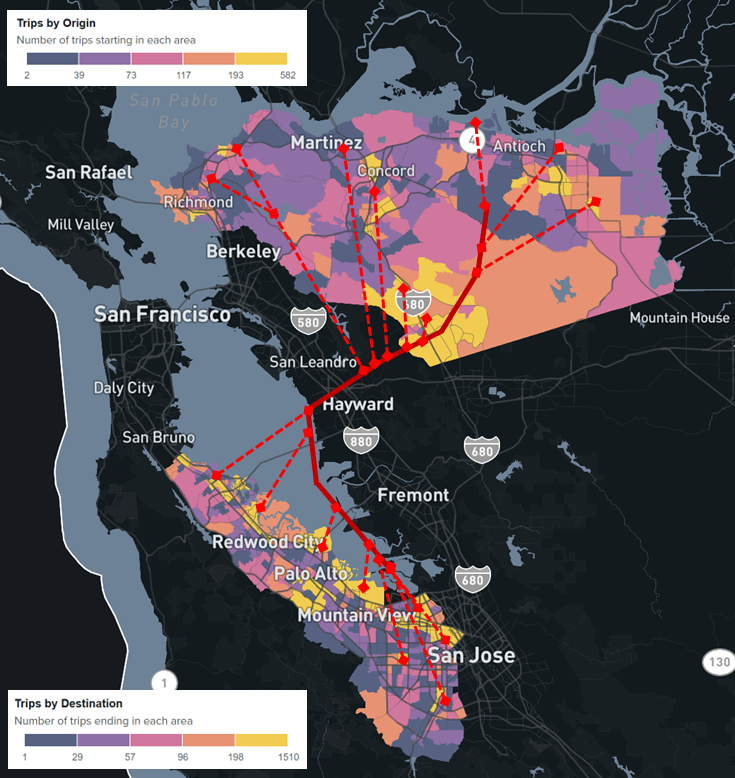}
\caption{Visualization of a high throughput UAM corridor concept in the Bay Area, connecting Contra Costa County and Silicon Valley. Map taken directly from Replica. Heatmap is divided into census tract and represents the absolute demand volume on a typical Thursday in Spring 2025. Vertiport locations were selected based on the closeness and intensity of the demand. \cite{rubi2025ucairlink, kam2025operational}}
\label{fig:network_design}
\end{figure}

\section{Experimentation}
\label{sec:experimentation}

\subsection{UAM Network and Model Parameters}
To conduct an experiment with our formulation, we model the UAM network in the Bay Area with a single high-throughput corridor connecting two aggregated vertiport clusters in the Bay Area, as shown in Fig. \ref{fig:network_design}. There is a 10-vertiport cluster in Contra Costa (CC) County, and an 8-vertiport cluster in Silicon Valley (SV), which captures the inter-region trip demand from nearby census tracts. In this paper, this corridor is referred to as the CC--SV corridor. 

Disaggregated travel demands between these clusters are sourced from Replica \cite{Replica2026} and treated as exogenous. To model the passenger flow from their origin to destination, we divided individual trips into a three-segment multimodal sequence: first-mile and last-mile by car or taxi, and middle-mile by UAM. Passengers' origin and destination areas are analyzed at census tract level, and average travel times between census tracts at a given time slot are used to compute passenger arrival times at each vertiport. Using simplified flight mechanics equations used in \cite{emin_evtol_2024}, the expected arrival times at the destination vertiport are computed. Each segment in the multi-modal trip sequence can therefore be represented by a travel time cost matrix for census tracts or inter-vertiport flights. This processed demand data represents the passenger-level demand for UAM operators, who will then compute and commit their intended UAM flight plans.

For simplicity, we aggregate the demand from ground travel data and use a fixed demand capture rate to compute the passenger arrival at vertiports as described above. We then use a rule-based dispatch policy to convert this to airspace demand, $F_{ijt}$. Specifically, operators will only depart if each vehicle is at least 75\% full (3 passengers out of 4 passengers), and passengers will spill after a fixed amount of waiting time. This is consistent with the existing simulation techniques for on-demand UAM operation \cite{emin_evtol_2024}. 

\begin{figure}[t!]
\centering
\includegraphics[width=\textwidth]{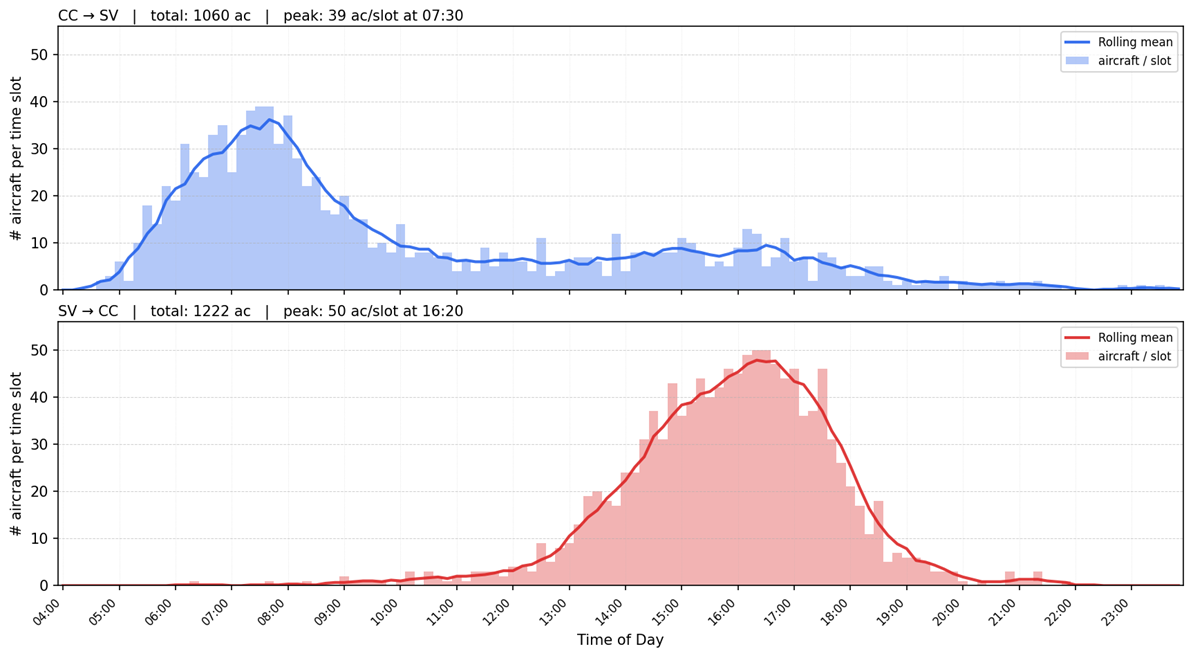}
\caption{Daily airspace-demand fluctuation for the CC--SV UAM corridor on a representative Thursday. Demand is computed at a 10-minute slot resolution from Replica ground-trip data with a UAM capture rate of $p_{c} = 0.3$, an aircraft seat capacity of 4, a 3-seat minimum dispatch load, and a maximum passenger wait of one slot (10~min) before spillage.}
\label{fig:airspace_demand}
\end{figure}

In our model, the intended UAM flight plans translate directly to the aircraft flow demand volume for the high-throughput UAM corridor, which we use to compute the expected ingress time to the airspace for each flight. Figure~\ref{fig:airspace_demand} shows daily demand fluctuation for the CC--SV UAM corridor. We aggregate this airspace demand over each time slot interval to solve a $T$-time step planning horizon optimization problem discussed in the Methodology section. We believe that to ensure the service level to meaningfully impact the urban transportation, where public transit is scarce, it is very effective to implement this supply side control by increasing the airspace capacity. This provides the fundamental design changes to the airspace necessary to produce high throughput and reduce unmet demands. 

The following parameter values were used for experimentation. The corridor admits $L_{ij} = 6$ physical lanes shared across both directions, each with per-slot throughput $K = 6$ aircraft. The planning horizon spans $[\,04{:}00,\, 24{:}00\,]$ in $\Delta t = 10$~min slots ($T = 120$ time steps), with flush time $\tau = 2$ slots. The market capture rate is fixed at $p_c = 0.30$ for the comparison study, with dispatch parameters $\mathrm{CAP} = 4$, $\mathrm{MIN\_LOAD} = 3$, and maximum passenger wait $W = 1$ slot. The objective weights used were $(C_u,\, C_s,\, C_w) = (1.0,\, 0.1,\, 0.01)$ which encode the priority order of fulfilling demand, limiting switching friction, and avoiding wasted capacity.

\subsection{Evaluation Metrics}
\label{sec:baselines}

We compare the dynamic policy against three baselines to assess whether its flush-time-aware switching and planning-horizon-wide allocation translate into measurable performance. Each baseline isolates one dimension of the decision space. All four policies produce a lane-allocation schedule $y_{ijt}$ over the same planning horizon and the same demand realization $F_{ijt}$, and are scored by a common evaluation routine so that observed differences are attributable to policy choice rather than measurement convention. The four policies vary in terms of two dimensions. The first is the timescale at which demand information enters the allocation decision. The second is whether the policy accounts for the flush time $\tau$ that physically constrains lane reversal.

Fixed 50/50 splits the available lanes equally between the two directions for the full horizon, $y_{ij,t} = \lfloor L_{ij}/2 \rfloor$ for all $t$. With our $L_{ij} = 6$ corridor, this is a three-on-three split. The baseline represents the maximally information-poor static configuration and serves as a lower bound against which any informed policy must improve. Fixed Asymmetric implements an operator-prescribed time-of-day schedule reflecting historical knowledge of the corridor's peak-direction pattern. The CC--SV schedule deployed here uses three blocks. In the morning window (04:00--12:00), five lanes are SV-bound and one is CC-bound. The afternoon (12:00--19:00) reverses this to one SV-bound and five CC-bound. The evening off-peak (19:00--24:00) operates with three lanes in each direction. The schedule is fixed at deployment time and does not respond to realized demand within an operating day. It represents the demand-awareness that a knowledgeable human dispatcher can embed into a static configuration on the basis of historical commute patterns. The policy switches lane direction only at the three scheduled block boundaries and does not plan around $\tau$ when doing so.

Greedy Reactive allocates lanes each slot in proportion to the instantaneous demand split, holding the previous slot's allocation when both directions have zero demand, with $y_{ij,t} = \mathrm{round}\!\left(L_{ij} \cdot F_{ij,t} / (F_{ij,t} + F_{ji,t})\right)$. The policy is purely myopic. It neither anticipates near-term demand changes nor accounts for the flush time that physically constrains lane reversal. This is the natural strawman for the dynamic policy, isolating what flush-aware planning adds over flush-blind reactivity. 

Dynamic is the proposed MILP formulation configured to optimize over the entire planning horizon ahead of execution. Each policy produces a schedule $y_{ijt}$ that is consumed by a common, policy-agnostic evaluation routine. For every directed edge $(i,j)$ and each slot $t$ the routine recovers the per-slot shortfall $s_{ijt} = \max(0,\, F_{ijt} - K \cdot y_{ijt})$, the per-slot waste $w_{ijt} = \max(0,\, K \cdot y_{ijt} - F_{ijt})$, the served aircraft $\min(F_{ijt},\, K \cdot y_{ijt})$, and the deactivation count $v_{ijt} = \max(0,\, y_{ij,t-1} - y_{ijt})$. We report four aggregate metrics across all directions and the full horizon. Total shortfall $\sum_{t,(i,j)} s_{ijt}$ in aircraft is the principal service-level indicator. Total waste $\sum_{t,(i,j)} w_{ijt}$ in aircraft-slots measures over-allocation. Total deactivations $\sum_{t,(i,j)} v_{ijt}$ measure operational friction. Mean lane utilization $\overline{F_{ijt} / (K \cdot y_{ijt})}$, averaged over slots with $y_{ijt} > 0$, measures how productively active capacity is used. Lastly, to quantify the service level improvement using the lane-based system, we use person-hours saved and average travel time per trip in the region of interest.

\section{Results}
\label{sec:results}

\begin{figure}[t!]
\centering
\includegraphics[width=\textwidth]{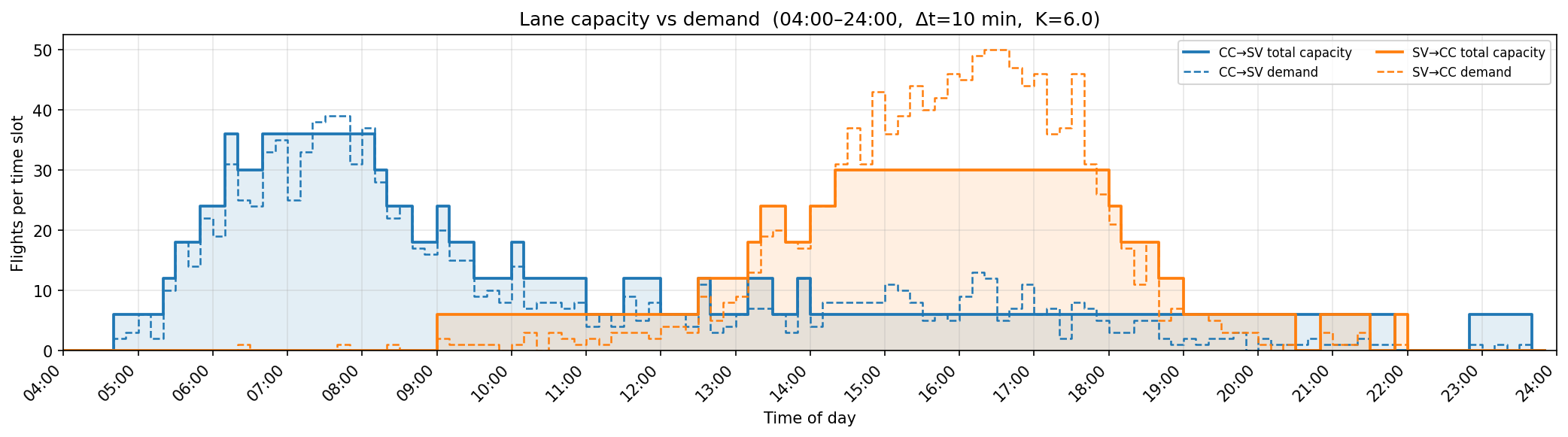}
\caption{Allocated corridor capacity (solid step function) under the dynamic policy, overlaid against the realized airspace demand $F_{ijt}$ (dashed) for both directions on a representative Thursday. The dynamic policy loads all lanes SV-bound during the morning commute peak. The policy transfers capacity CC-bound to follow the long afternoon return peak, then releases lanes to the idle pool as demand in both directions subsides in the evening.}
\label{fig:dynamic_capacity_demand}
\end{figure}

Figure~\ref{fig:dynamic_capacity_demand} shows the dynamic policy's allocated capacity overlaid against the realized airspace demand. The optimizer recovers the canonical commute pattern. In the morning it allocates all six available lanes to the SV-bound direction, matching a sharp peak that temporarily reaches 39 aircraft per slot, then gradually shifts capacity to the CC-bound direction as morning demand subsides into the afternoon. From 14:20 to 18:00 it holds a five-lane CC-bound allocation through the long afternoon return peak. During this peak, maximum demand of approximately 50 aircraft per slot exceeds the physical maximum throughput capacity of $K \cdot L_{ij} = 36$ aircraft per slot even when all lanes are loaded into a single direction, so some shortfall in this window is unavoidable for any policy. After 19:00 the policy drops sharply to one lane per direction and holds that allocation through the late evening, decaying to zero in the final slots. The unallocated lanes during off-peak hours are returned to the idle pool, recovering the supply-side latitude that motivated the inequality form of Eq.~\eqref{eq:physical_capacity}.

\begin{figure}[b!]
\centering
\includegraphics[width=\textwidth]{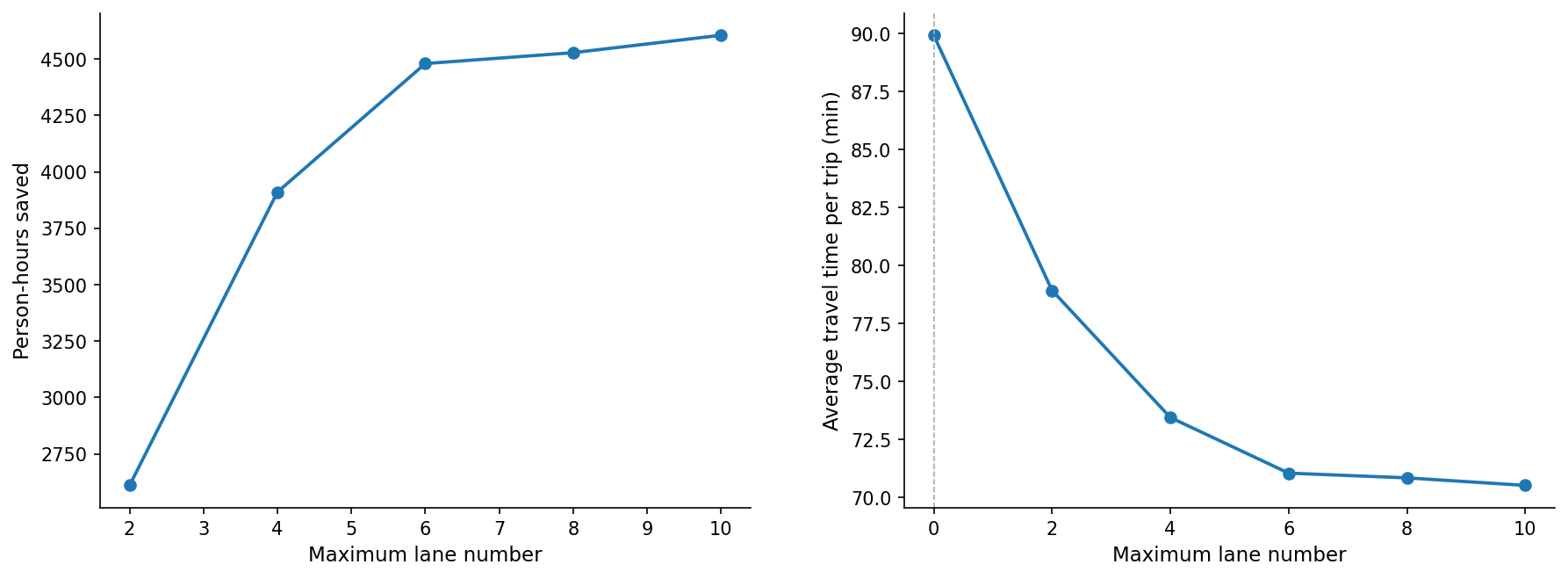}
\caption{Population-level travel-time impact of maximum lane number at $30\%$ market capture, $\tau = 2$, $K = 6$. Left: total person-hours of travel time saved per day relative to the no-UAM baseline. Right: mean travel time per trip averaged across the full commuting population, with the no-UAM reference (around 90 minutes) shown as a dashed line at $L = 0$.}
\label{fig:pmt_summary}
\end{figure}

Figure~\ref{fig:pmt_summary} sweeps maximum lane number $L_{ij}$ from 2 to 10 lanes under the dynamic policy and reports the resulting impact on the entire commuting population. This includes all passengers who traveled by UAM and driving. Both metrics exhibit clear diminishing returns. The first two lanes recover roughly 2{,}610 person-hours of travel time per day; doubling to $L = 4$ adds another 1{,}300 person-hours, and the next two doublings contribute 570 and 50 person-hours respectively. By $L = 8$ the curve has flattened substantially, and the marginal benefit from $L = 8$ to $L = 10$ is only about 80 additional person-hours. Population mean per-trip travel time drops from 89.9 minutes under the no-UAM baseline to 70.5 minutes at $L = 10$, a $21.6\%$ reduction.

The asymptote is set not by $L_{ij}$ but by the upstream dispatch policy. Of the 14{,}241 total addressable passengers each day, 5{,}113 are dropped at the origin vertiport because they exceed the maximum-wait deadline before being pooled into a full aircraft --- the dispatch policy's $\mathrm{MIN\_LOAD} = 3$, $\mathrm{CAP} = 4$ operating point favors load-factor discipline over service universality. In reality, this number can further decline due to other capacity constraints such as the number of gates and final approach and take off (FATO) area limits at vertiports. These passengers spill to driving regardless of how much maximum lane number is allowed. As a result, the population-level saving flattens at $L_{ij}=6$: beyond that, additional lanes cannot serve passengers the dispatch policy has already dropped.

This decomposition matters for siting and scaling decisions. Once $L_{ij}$
is past the knee of the curve (around $L = 6$ to $L = 8$ at this demand level),
further investment in lane count produces marginal social return; throughput gains
must instead come from relaxing the dispatch policy or from expanding vertiport-side throughput. The contribution of dynamic lane allocation in this regime is therefore best framed as \emph{capacity-efficient} --- delivering close to the achievable
benefit at moderate $L$.

\subsection{Comparison with Baselines}

\begin{figure}[b!]
\centering
\includegraphics[width=\textwidth]{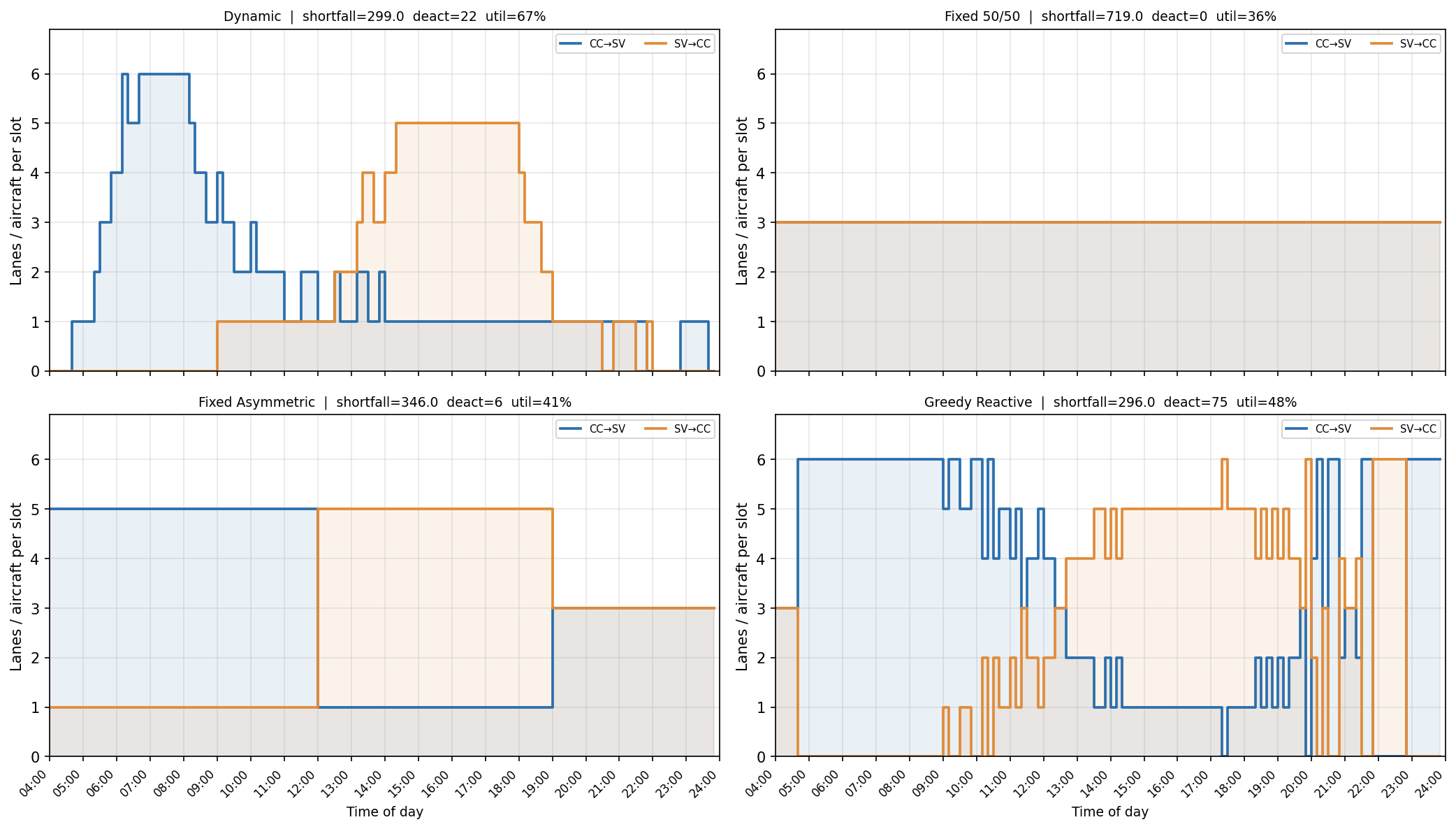}
\caption{Lane allocation schedule over time for each of the four policies on the same airspace demand realization. The dynamic policy (top left) produces a piecewise-stable schedule that follows
the AM and PM peaks; Fixed 50/50 (top right) holds three lanes in each direction throughout; Fixed Asymmetric (bottom left) tracks the
operator's three-block schedule; Greedy Reactive (bottom right) reconfigures repeatedly in response to instantaneous demand fluctuation.}
\label{fig:comparison_timeseries}
\end{figure}

Four lane management policies were compared on a single representative Thursday's demand realization with fixed maximum lane number $L_{ij} = 6$, lane
throughput $K = 6$ aircraft per slot, flush time $\tau = 2$ slots ($20$~min), and capture rate $p_{c} = 0.3$. All four policies managed the same airspace demand $F_{ijt}$ and are scored by the common evaluation metrics outlined in Sec.~\ref{sec:baselines}. Figure~\ref{fig:comparison_timeseries} compares the dynamic allocation pattern against the three baselines on the same demand realization. 

Notably, the greedy policy tracks demand at the slot level but flips lane direction repeatedly across the horizon. Every short-duration excursion in demand causes a re-allocation, and the policy cannot settle into a stable block structure even during sustained peaks. The dynamic policy by contrast produces a piecewise-stable schedule whose transitions align with the demand-regime changes.

\begin{figure}[t!]
\centering
\includegraphics[width=\textwidth]{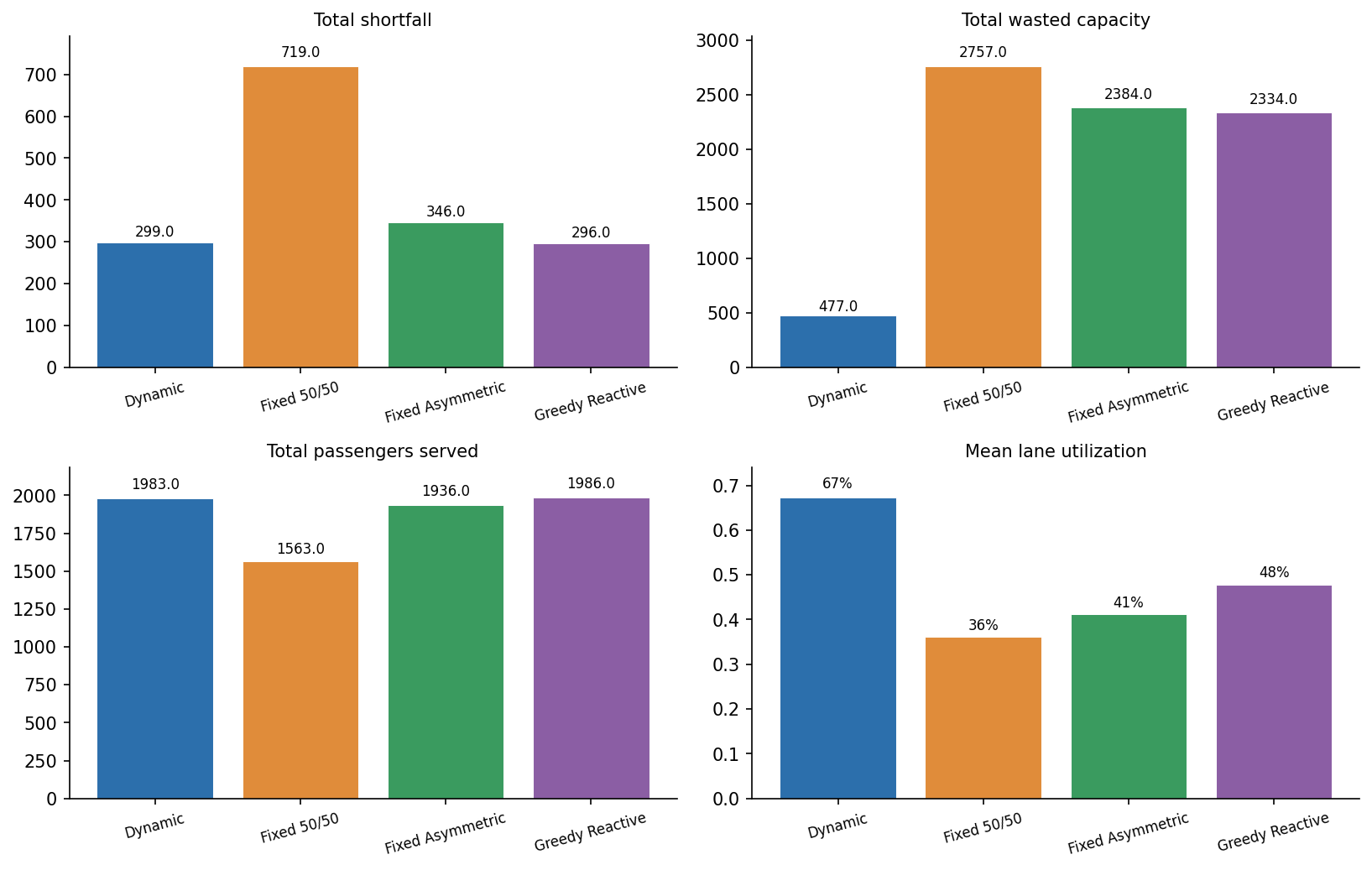}
\caption{Aggregate performance metrics for the four policies on a
representative weekday at $L_{ij} = 6$, $p_{c} = 0.3$.
The dynamic policy attains the highest mean lane utilization
($67\%$), reduces wasted capacity by approximately $5\times$
relative to the three baselines, and delivers shortfall
($299$ aircraft) essentially tied with Greedy Reactive at a small
fraction of the operational reconfigurations.}
\label{fig:comparison_metrics_bar}
\end{figure}

Figure~\ref{fig:comparison_metrics_bar} reports the four performance metrics aggregated across the planning horizon. The dynamic policy delivers a shortfall of $299$ aircraft (a shortfall rate of $13.1\%$), essentially tied with Greedy Reactive ($296$ aircraft, $13.0\%$), beating Fixed Asymmetric ($346$, $15.2\%$) by $2.1$ percentage points and Fixed 50/50 ($719$, $31.5\%$) by $18.4$ percentage points. The shortfall improvement against the best baselines is small in absolute terms because the better-informed policies all push against the same physical-capacity ceiling during the evening peak; the dynamic policy's contribution beyond that shared ceiling is more visible in the secondary metrics. Total wasted capacity drops by approximately $5\times$ from $2{,}334$--$2{,}757$ aircraft-slots under the baselines to $477$
under the dynamic policy, and mean lane utilization rises from $36$--$48\%$ to $67\%$. The static baselines waste capacity by holding lanes open in the wrong direction during peaks; the greedy baseline wastes capacity by spreading allocation across both directions and absorbing dead time from frequent transitions.

The deactivation count exposes the contribution most directly attributable to the formulation's flush-time-aware structure. The dynamic policy reconfigures the corridor $22$ times across the $120$-slot horizon; the greedy reactive baseline reconfigures $75$ times --- $3.4\times$ as often --- to deliver effectively identical service ($1{,}983$ versus $1{,}986$ aircraft served, a $0.15\%$ gap). The fixed asymmetric schedule reconfigures only $6$ times by construction but pays for that simplicity in $5\times$ higher wasted capacity. The dynamic policy occupies a region of the trade-off space that none of the baselines reaches. Low operational friction at the same service level as the most reactive baseline, and substantially better airspace utilization than any static one.

The two demand-aware baselines are interesting to contrast directly. Fixed Asymmetric and Greedy Reactive achieve broadly similar service levels --- $1{,}936$ versus $1{,}986$ aircraft served, a $2.6\%$ gap in Greedy's favor --- with shortfalls of $346$ and $296$ respectively. The operator's static schedule, designed offline from prior knowledge of the corridor's commute pattern, captures most of the first-order demand structure that the greedy policy chases reactively. They differ sharply, however, in that Fixed Asymmetric requires only $6$ scheduled changes per day, as opposed to $75$ for Greedy Reactive, more than an order-of-magnitude more reconfigurations for a marginal service gain. This suggests that for corridors with predictable commute-style demand, the marginal value of online reactivity is small relative to a well-designed static schedule matched to the AM/PM peak directionality. The dynamic policy follows short-term demand changes without switching lanes every time. This balance comes from two parts of our model working together: the flush-time constraint (Eq.~\eqref{eq:physical_capacity}) and the switching penalty in the objective. They let the optimizer track demand closely but only switch when the benefit outweighs the cost.

The dynamic policy solves to optimality in $0.043$ seconds on the $120$-slot horizon, well within any operational planning timescale. The formulation is therefore suitable both for pre-tactical day-of-operations planning and for rolling-horizon re-solves under updated demand forecasts during the day.

\subsection{Sensitivity Analysis}

Next, in order to understand how the dynamic policy's performance depends on the two structural parameters, we conducted a parameter sweep study. For this, we fix the flush time $\tau = 2$ and vary the market capture rate $p_{c}$ and the maximum lane numbers $L_{ij}$ to see their effect. Specifically, we sweep over $L_{ij} \in \{2, 4, 6, 8, 10\}$ and $p_{c} \in \{0.1, 0.2, \dots, 0.8\}$. The 40 resulting combinatorial configurations are solved through the same model pipeline.

\begin{figure}[hbt!]
\centering
\includegraphics[width=\textwidth]{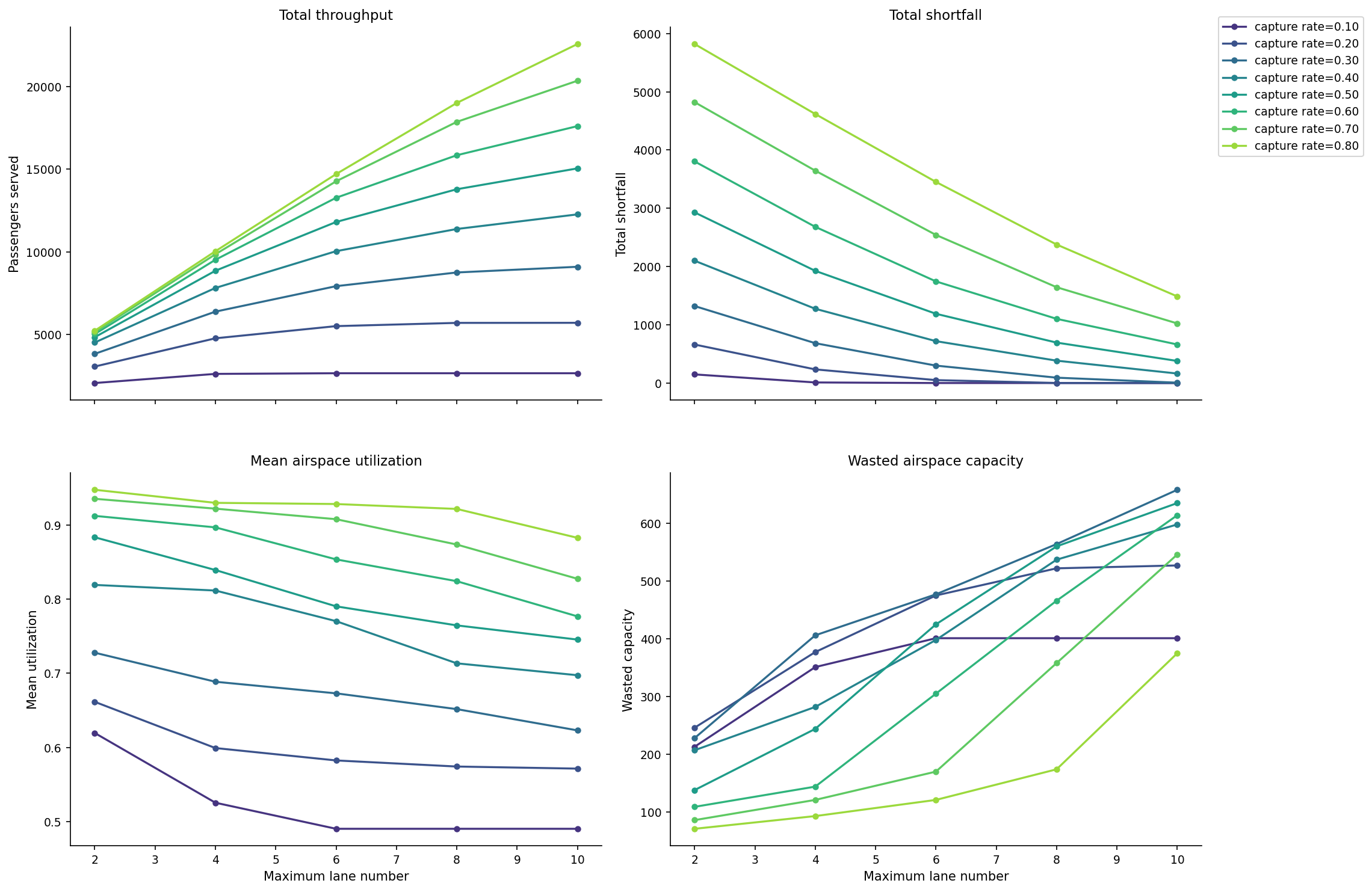}
\caption{Sensitivity of dynamic-policy performance to maximum lane number $L_{ij}$ and UAM capture rate $p_{c}$, across 40 sweep configurations. Each curve fixes a capture rate and varies $L_{ij}$ along the horizontal axis. The four panels jointly identify the operating regime of the corridor. All metrics scale with the maximum lane number until demand saturates, revealing diminishing return on increasing the maximum lane number.}
\label{fig:sweep_study}
\end{figure}

Throughput (top left) scales nearly linearly with maximum lane number at high capture rates and saturates at low capture rates. At $p_{c} = 0.1$, throughput rises from 2{,}076 served passengers at $L_{ij} = 2$ to a ceiling of 2{,}668 at $L_{ij} = 6$ and remains flat at higher lane numbers. In this case, the corridor is demand-limited, and additional lanes generate no additional service. At
$p_{c} = 0.8$, throughput grows nearly linearly from 5{,}236 at $L_{ij} = 2$ to 22{,}572 at $L_{ij} = 10$, gaining roughly 4{,}000--4{,}500 served passengers per pair of lanes added and not yet plateauing at the right edge of the sweep --- the corridor remains capacity-limited even at ten lanes when adoption is high.

Zero shortfall (the number of flights that were submitted but rejected due to airspace capacity limit) is reached at $L_{ij} = 6$ for $p_{c} = 0.1$ and at $L_{ij} = 10$ for $p_{c} = 0.2$, but no $(L, p)$ combination in the swept range drives shortfall to zero once the capture rate exceeds 0.3. The shortfall surface is monotonically decreasing in $L_{ij}$ and monotonically increasing in $p_{c}$, with diminishing returns to additional lanes. This concave pattern informs the corridor right-sizing decision: the operator selects the smallest $L_{ij}$ for which shortfall falls below an acceptable service threshold given the expected market capture rate. As for mean utilization (bottom left), the corridor operates well below half of its provisioned active capacity at low capture rate and high lane count. At high capture rate the active lanes run close to 90\% utilization even at $L_{ij} = 10$ ($88\%$ at $p_{c} = 0.8$). The utilization also quantifies the supply-side cost of over-allocation. A corridor sized more than its captured demand results in operational overhead on lanes that are open but underused.

Wasted capacity (bottom right) generally rises with $L_{ij}$, since more provisioned lanes create more opportunities for capacity to go unused. However, at the lowest capture rate, the curve flattens as maximum lane number $L_{ij}$ grows. At $p_{c} = 0.1$, wasted capacity reaches 401 aircraft-slots at $L_{ij} = 6$ and stays at exactly 401 for $L_{ij} = 8$ and $L_{ij} = 10$. The reason is that when demand is too low to justify additional active lanes, the optimizer leaves them in the idle pool. This is the result of Eq.~\eqref{eq:physical_capacity} which releases excess capacity to the idle pool. It confirms that adding lanes beyond the demand yields no additional throughput and no additional waste. This has a direct implication for capacity planning. Adding more lanes to a corridor immediately results in no additional wasted capacity after some point, since the optimizer leaves surplus capacity in the idle pool. Lanes provisioned ahead of demand growth therefore can simply be available for reallocation as the captured market matures, or for ad hoc reassignment to other flight purposes such as emergency transport or cargo missions without imposing operational overhead in the meantime. The sensitivity analysis reveals some important benefits and limitations of airspace management. There is a diminishing return on increasing the maximum lane number at a given demand level. Hence, increasing the maximum lane numbers does not increase the throughput or prevent shortfall completely due to airspace capacity limit. Moreover, at a fixed lane number, throughput becomes airspace-limited as market share grows, with throughput volume failing to scale linearly at the higher capture rates. 


\subsection{Application to Multimodal Transportation}


The formulation presented here is grounded in the Contra Costa County--Silicon Valley corridor, but the framing it enables is broader. The same approach applies to any door-to-door multimodal trip where an underserved or geographically separated region has insufficient ground-based connectivity to a major employment or economic center, and where eVTOL service can plausibly fill the middle-mile gap. Several candidate corridors in the Bay Area fit this profile. Hollister and the broader San Benito County area are tied to Silicon Valley employment but separated by Pacheco Pass and chronic US-101 congestion. The Monterey Bay region (e.g., Monterey, Salinas, Watsonville) lies at a similar distance from both Silicon Valley and San Francisco, with only a single freeway corridor connecting it to either. In each case the mobility challenge mirrors our case study. The demand is asymmetric and peak-directional, the corridor lane budget is finite, and the dynamic allocation problem is to translate that budget into responsive directional capacity.

For transportation planners, the formulation provides a tool for capacity right-sizing before infrastructure is committed. The sensitivity sweep in this section reveals the diminishing-returns regime explicitly, allowing planners to identify the smallest lane budget that satisfies a target service threshold for an expected adoption profile. Pairing this with vertiport siting analyses and ground-side transit integration plans produces an end-to-end design path that does not rely on aircraft availability projections alone. For ATM and PSU service providers, the formulation produces an executable lane-allocation policy at sub-second latency, well within the operational planning horizon for day-of-operations decisions and rolling re-solves. As COPs are defined under the FAA UAM ConOps, formulations like this one are candidates for inclusion in PSU service portfolios. They provide a configuration-layer decision rule that operates upstream of, and complementary to, the scheduling-layer services already in development.

The policy and community-engagement implications matter most in the near term. Vertiport siting and corridor design are local-government decisions with direct community impact. Demonstrating that dynamic corridor capacity can serve commute patterns from underserved or geographically separated communities without requiring full saturation of either direction strengthens the equity case for AAM service. Community outreach efforts, including transparent communication about flight density, noise envelopes, and service hours, are essential to securing the public trust required for vertiport siting and corridor operation. Our results also suggest that off-peak idle-lane capacity can be earmarked for non-commute public benefit purposes such as emergency medical transport or disaster response staging, which strengthens the public-good case that policymakers and community advocates will want to see. For instance, personal transportation is greatly reduced at night. But such time slots may require increased capacity for emergency vehicles or air-based cargo deliveries. Together, these considerations position the dynamic lane allocation problem as one of several enabling decisions for the near-term scaling of AAM within multimodal transportation systems. The formulation does not depend on full corridor saturation, does not require a single deployer to hold both configuration and scheduling authority, and supports rolling re-solves under updated forecasts. It can therefore be deployed alongside the regulatory and community-engagement work that real-world AAM rollout will require.

\section{Conclusion}

This article presents a discrete-time MILP for dynamic directional lane allocation within a UAM corridor. We show that our formulation cuts wasted capacity by approximately $5\times$ relative to informed static and reactive baselines and lifts mean lane utilization from 36--48\% to 67\%. These supply-side improvements translate into door-to-door multimodal service quality, with the commuting population seeing up to a 21.6\% reduction in mean per-trip travel time from people's final origin-to-destination trip sequences. The design degree of freedom this exploits is intrinsic to airspace as an operational medium. Airspace can be reallocated without the physical infrastructure changes that limits similar system on the ground. This framing can be generalized beyond a single corridor. Any door-to-door multimodal use case where a geographically separated or underserved region has insufficient ground-based connectivity to a major economic center exhibits the same structural pattern of asymmetric peak-direction demand on a finite maximum lane budget. The formulation provides a tool for capacity right-sizing and operational planning in each such corridor, supporting the configuration-layer decisions that planners, PSU serviders, and policymakers will need to coordinate as AAM scales.

\subsection{Limitations and Future Works}

Several limitations of the present study constrain its immediate deployment. First, the model treats the corridor network as a fixed geometric input. The spatial path of each corridor through 3D urban airspace is given, and we optimize only the directional allocation of lanes within each corridor. In practice, corridor geometry itself is a decision variable for the airspace manager, particularly under time-varying weather that may render a nominal corridor path unsafe and force a reroute through alternative airspace at potentially lower capacity. Second, we assume a single deployer with authority to configure lane directions. The UAM ConOps does not yet assign this authority, and intra-corridor configuration in a federated PSU environment will likely require negotiated consensus among multiple operators, with attendant equity considerations for how directional capacity is allocated to demand from competing service providers.

Future work includes integrating corridor geometry selection with directional capacity allocation in a unified or hierarchically decomposed optimization, which may accommodate weather- and energy-informed path selection. Extending the formulation to support prioritized demand classes and a richer set of preemption rules would enable short-horizon re-allocations triggered by medical or emergency missions, sudden demand surges, or temporary closure of adjacent corridors. Finally, the formulation can be extended to the multi-PSU setting, where directional capacity and flight slots must be allocated across competing operators in a way that all federated participants can accept.


\section{Acknowledgments}

This work is funded by the Center for Information Technology and Research in the Interest of Society (CITRIS) through the CITRIS Aviation Prize 2024--2025. Andrew Park was supported by the California Partners for Advanced Transportation Technology (PATH) and Jordan Kam was supported by NASA's California Space Grant Consortium under Award \#80NSSP20M0099. The authors would like to thank Professor Mark Hansen for his insights in problem formulation and demand modeling. Lastly, we thank Dr.~Husni Idris from NASA Ames Research Center for discussions on the Contra Costa case study.

\bibliography{reference}

\end{document}